\documentclass[12pt,english]{article}
\usepackage[T1]{fontenc}
\usepackage[latin9]{inputenc}
\usepackage{amsmath}
\usepackage{amssymb}
\usepackage{graphicx}
\usepackage{esint}

\makeatletter

\usepackage{amsfonts} 
\newcommand{\eq}{\begin{equation}}
\newcommand{\eqn}[1]{\label{#1}\end{equation}}
\newcommand{\eea}{\end{eqnarray}}
\newcommand{\eqa}{\begin{eqnarray}}
\newcommand{\eqan}[1]{\label{#1}\end{eqnarray}}
\newcommand{\ba}{\begin{array}}
\newcommand{\ea}{\end{array}}
\newcommand{\eqac}{\begin{equation}\begin{array}{rcl}}
\newcommand{\eqacn}[1]{\end{array}\label{#1}\end{equation}}

\usepackage{babel}

\makeatother

\usepackage{babel}
\begin{document}

\title{\textbf{Complex Chern-Simons and the Gribov Scenario for Confinement}\\
 \textbf{ }}

\author{M. M. Amaral$^{a,b}$\footnote{email: mramaciel@gmail.com} ,
V. E. R. Lemes$^{a}$\footnote{email: vitor@dft.if.uerj.br},
O. S. Ventura$^{c}$\footnote{email: ozemar.ventura@cefet-rj.br} ,
L. C. Q. Vilar$^{a}$\footnote{email: lcqvilar@gmail.com}  \\
\small \em $^a$Instituto de F\'\i sica, Universidade do Estado do Rio de
Janeiro,\\
\small \em Rua S\~{a}o Francisco Xavier 524, Maracan\~{a}, Rio de Janeiro - RJ,
20550-013, Brazil\\
\small \em $^b$Institute for Gravitation and the Cosmos $\&$ Physics Department,\\
\small \em The Pennsylvania State University, University Park, PA 16802, USA\\
\small \em $^c$Centro Federal de Educa\c{c}\~ao Tecnol\'ogica do Rio de
Janeiro\\
\small\em Av.Maracan\~a 249, 20271-110, Rio de Janeiro - RJ, Brazil}

\maketitle
\vspace{-1cm}

\begin{abstract}


We show that BLG-ABJM type of theories, discovered in the context of the AdS/CFT correspondence, generate gauge propagators with the complex pole structure prescribed by the Gribov scenario for confinement, which was developed in context of Yang-Mills theories. This structure, known as \textit{i-particles} in Gribov-Zwanziger theories, effectively allows the definition of composite operators with a positive K\"{a}ll\'{e}n-Lehmann spectral representation for their two-point functions. Therefore these operators satisfy the criteria to describe glue-ball condensates. We calculate the (first order) contribution to the two-point function  of the gauge invariant condensate in the ABJM framework. We show that the correlator can be interpreted as a physical composite particle state in terms of the K\"{a}ll\'{e}n-Lehmann representation.

\end{abstract}
\setcounter{page}{0}\thispagestyle{empty}

\vfill{}
 \newpage{}\ \makeatother

\section{Introduction}

Three-dimensional gauge theory is an interesting laboratory for many studies in non-perturbative aspects of gauge field theories such as color confinement \cite{Li1995,Ivanov1998,Mathieu:2008me} or topological properties such as obtained from Chern-Simons action \cite{Witten1989,Witten:2010cx,Witten:2010zr}. Also, three-dimensional Yang-Mills theory has local degrees of freedom and the coupling constant has dimension of mass. This properties indicates that this theory can be seen as an approximation for the high temperature phase of QCD
with the mass gap in the role of the magnetic mass \cite{Ivanov1998}. In particular one of the mechanisms to study color confinement comes from the analysis of Gribov copies \cite{Gribov}, known generally as Gribov ambiguity, with special emphasis on the Gribov-Zwanziger model (GZ) \cite{Zwanziger1,Zwanziger2,Zwanziger25,Zwanziger3}  and its refined version (RGZ) \cite{rgzmodel}.
One of the Gribov mechanism properties is that it generates propagators for gauge fields with complex poles, known as \textit{i-particles}\cite{Sorella:2010it,Capri:2010pg}. The \textit{i-particles} can not be identified with the propagation of simple massive particles but they provide the possibility to obtain condensates that behave like massive particles. This property is interpreted as confinement 
and known generally as Gribov-Zwanziger scenario. The Gribov ambiguity is a general characteristic of the quantization of Yang-Mills theories and a general property of all local covariant renormalizable gauge fixing procedures \cite{singer}. 

It is important to emphasize here that the Gribov ambiguity is intrinsically linked to the Morse theory as discussed by van Baal \cite{Baal1992}. His work begins by interpreting the Gribov copies in its variational form as a problem in Morse theory\footnote{See also \cite{labast} for a 
previous use of Morse theory on topological quantum field 
theories and its relation to the Gribov ambiguity.}. Into simple terms, Morse theory searches for a 
characterization of topological invariants of any given manifold by the study of the critical points 
of functions defined on it \cite{milnor,matsumoto}, like the Hilbert norm $I_S={Tr\int_{M} {\tilde{A}_i}^2}$, which can be used to 
study the Gribov mechanism \cite{Guimaraes2016,Capri2013,Vilar2011}. 
Morse theory is also important in the analytic continuation of three-dimensional Chern-
Simons gauge theory away from integer values of the usual coupling parameter $k$. This analytic continuation can be carried out
by generalizing the usual integration cycle in the Feynman path integral \cite{Witten:2010cx,Witten:2010zr}. Morse theory gives a natural framework for describing the appropriate integration cycles. 

Another aspect of three-dimensional gauge theories, which we will consider in this paper arise in the context of the AdS/CFT 
correspondence in the form of Bagger-Lambert-Gustavsson (BLG) theory \cite{Bagger, Gustavsson:2007vu} and Aharony-
Bergman-Jafferis-Maldacena (ABJM) theory \cite{Aharony:2008ug}. The BLG and ABJM theory are good candidates for 
the dual superconformal field theory in the correspondence. There is a usual approach to confinement in the AdS/CFT correspondence that study the dual gravity description \cite{Mathieu:2008me,Brower2000,Colangelo2007}. In this
paper we study the gauge and scalar field sectors of the dual superconformal field theory.
The BLG, or at least ABJM, theory in the case of $U(1)_{k}\times U(1)_{-k}$ 
can be cast in the form of a complex holomorphic theory and this structure appears to be fundamental in every gauge theory in which the infrared sector is under study. 

The main objective of this paper is to present a study of some properties of the Gribov-Zwanziger scenario that arise in this complex theory, with particular emphasis on the relation between the \textit{i-particles} pole structure, complex gauge theory and observables that admits K\"all\'{e}n-Lehmann spectral representation.
We investigate the Supersymmetric Chern Simons (SCS) theory (N = 1, D = 3), with superfields formalism. SCS coupled with the Super Yang Mills (SYM) action is introduced to present the structure of the Gribov ambiguity in three dimensions. The same SCS action is the base of BLG-ABJM models. We propose that a complex gauge structure can be related to the Gribov scenario by showing that the \textit{i-particles} can be obtained from complex Chern Simons (CS) theory due to a symmetry breaking mechanism in the BLG-ABJM model \cite{Mukhi2008,Mauri2008,Mukhi2011}. Of course, the symmetry breaking mechanism can not be understand in the context of a minimum of a scalar potential due to the non positivity of the action when written in terms of real gauge fields. The symmetry breaking mechanism should be understood as a critical point \cite{Mukhi2008} that permits the access to infrared properties due to the introduction of a scale and the fact that the theory is not topological anymore. 

The paper is organized as follows: in Section 2, the problem of Gribov ambiguity is presented for the Super Yang Mills - Chern Simons (SYM-CS) theory and the \textit{i-particles} pole structure is introduced. Their relation to complex fields is discussed. In section 3 the Complex generalization of CS is presented and a complex scalar field is introduced in context of the BLG-ABJM model. The symmetry breaking mechanism is presented and the \textit{i-particles} structure as well as the construction of gauge invariant composite operators is obtained.

\section{SYM-CS theory with Gauge fixing and Gribov ambiguity (Superfield, N = 1, D = 3)}

In three-dimensional Minkowski space-time the Lorentz group is $SL(2,R)$ and the corresponding fundamental representation
acts on a two components real (Majorana) spinor. In the case of Euclidean $D = 3$, the two components spinor shall be transformed under $SO(3)$ and as is well known 
\cite{Kugo:1982bn,McKeon:2000qm,McKeon:2001pm} one can not have the usual
Majorana condition. This is the same situation we have in $D = 4$ \cite{Amaral:2013uya}.
So we take the approach of generalizing the concept
of complex conjugation of Grassmann algebra \cite{Wetterich:2010ni}.
In this way, it is possible to have the condition of Majorana in the Euclidean space and thus the same field content of the $N=1$ case. 
The notations and conventions are in Appendix A. Let
us take the Euclidean version of the superspace action of SYM-CS \cite{RuizRuiz1}:
\begin{equation}
S_{SYMCS}=S_{SYM}+S_{SCS},\label{eq:actionSYMCS}
\end{equation}
with,
\begin{equation}
S_{SYM}=\frac{1}{2}\int d^{3}xd^{2}\theta W^{a\alpha}W_{\alpha}^{a},\label{eq:actionSYM}
\end{equation}
and
\begin{equation}
S_{SCS}=im\int d^{3}xd^{2}\theta\left[(D^{\alpha}\Gamma^{a\beta})(D_{\beta}\Gamma_{\alpha}^{a})+\frac{2}{3}igf^{abc}\Gamma^{a\alpha}\Gamma^{b\beta}(D_{\beta}\Gamma_{\alpha}^{c})-\frac{1}{6}g^{2}f^{abc}f^{cde}\Gamma^{a\alpha}\Gamma^{b\beta}\Gamma_{\alpha}^{d}\Gamma_{\beta}^{e}\right].\label{eq:actionSCS}
\end{equation}
The field strength is given by
\begin{equation}
W_{\alpha}^{a}=D^{\beta}D_{\alpha}\Gamma_{\beta}^{a}+igf^{abc}\Gamma^{b\beta}D_{\beta}\Gamma_{\alpha}^{c}-\frac{1}{3}g^{2}f^{abc}f^{cde}\Gamma^{b\beta}\Gamma_{\beta}^{d}\Gamma_{\alpha}^{e},\label{woperator}
\end{equation}
and superspace derivative
\begin{equation}
D_{\alpha}=\frac{\partial}{\partial\theta^{\alpha}}+i\sigma_{\alpha}^{\mu\gamma}\varepsilon_{\gamma\beta}\theta^{\beta}\partial_{\mu}.
\end{equation}

The supermultiplet of gauge fields in Wess-Zumino gauge is
\begin{align}
\Gamma_{\alpha}^{a}(x,\theta) & =i\sigma_{\alpha}^{\mu\gamma}\varepsilon_{\gamma\beta}\theta^{\beta}A_{\mu}^{a}(x)+i\theta^{2}\lambda_{\alpha}^{a}(x).\label{eq:spinorsuperfieldcomponent}
\end{align}
They belong to the adjoint representation of the gauge group $SU(N)$. 

The classical action for SYM-CS theory, $S_{SYMCS}$, remains invariant
under the following gauge transformation 
\begin{equation}
\delta_{\Lambda}\Gamma_{\alpha}^{a}=(\nabla_{\alpha}\Lambda)^{a},\label{eq:gaugetransformation}
\end{equation}
with superspace covariant derivative given by
\begin{equation}
\nabla_{\alpha}^{ab}=\delta^{ab}D_{\alpha}+gf^{acb}\Gamma_{\alpha}^{c}.\label{superspacecovariantderivative}
\end{equation}

\subsection{Gauge-fixing}

In order to quantize the theory correctly we have to fix the gauge
and we can do it covariantly using the usual procedure of Faddeev-Popov
(FP) on the supersymmetric Landau gauge. We implement the conditions
$D^{\alpha}\Gamma_{\alpha}^{a}=0$. 
Following these procedure we end with the action of gauge
fixing
\begin{equation}
S_{gf}=\frac{1}{4}s\{\int d^{3}xd^{2}\theta(c'^{a}D^{\alpha}\Gamma_{\alpha}^{a})\},\label{eq:actiongaugefix}
\end{equation}
where the Faddeev-Popov ghost fields will be scalar superfield.
The fields $c'^{a}$ and $c^{a}$ are the antighost and the ghost respectively.
And $s$ is the BRST nilpotent operator ($s^{2}=0)$.

The total action $S=S_{SYMCS}+S_{gf}$ is invariant under the BRST
transformations \cite{RuizRuiz1}:
\begin{align}
s\Gamma_{\alpha}^{a} & =(\nabla_{\alpha}c)^{a}\nonumber \\
sc^{a} & =-\frac{1}{2}gf_{abc}c^{b}c^{c}\nonumber \\
sc'^{a} & =b^{a}\nonumber \\
sb^{a} & =0.\label{eq:brst}
\end{align}

Using this gauge fixing the massive gauge propagator for SYM-CS is:
\begin{equation}
<\Gamma_{\alpha}^{a}(1)\Gamma_{\beta}^{b}(2)>=\frac{\delta^{ab}}{\partial^{2}(-\partial^{2}+m^{2})}(D^{2}-im)D_{\beta}D_{\alpha}\delta^{2}(\theta_{1}-\theta_{2})\delta^{3}(x_{1}-x_{2}).\label{eq:SYMCSpropagator}
\end{equation}

\subsubsection{SYM-SC and Gribov Ambiguity}
 
The problem of Gribov ambiguity is a general property of all local covariant renormalizable gauge fixing  \cite{singer}.
It is straightforward to note that the Landau gauge condition is not ideal. If we consider two equivalents superfield, $\Gamma_{\alpha}^{a}$ and $\Gamma_{\alpha}^{a'}$, connected by a gauge transformation (\ref{eq:gaugetransformation}), and if both satisfy the same condition of the Landau gauge, $D^{\alpha}\Gamma_{\alpha}^{a}=0$ and $D^{\alpha}\Gamma_{\alpha}^{a'}=0$, we have
\begin{equation}
D^{\alpha}(\nabla_{\alpha}\Lambda)^{a}=0.\label{FPoperatorSusyini}
\end{equation}
Therefore, the existence of infinitesimal copies, even after FP quantization is related to the presence of the zero modes of  the operator above.
This suggests that we should restrict the functional integration to a region free of zero modes, 
and free of gauge superfields copies.
To do this we would like to study the operator (\ref{FPoperatorSusyini}) in terms of the eigenvalues and eigenvectors equation, which is not immediately possible since the equation $D^{\alpha}(\nabla_{\alpha}\Lambda)^{a}=\lambda\Lambda$ is not an eigenvalue equation.
This can be seen in field components where the equation relate different components of the superfiel $\Lambda$. 
This indicates that the correct operator, where one can study the  zero modes problem, and thus define the problem of Gribov ambiguity, is
\begin{equation}
{\cal O}^{ab}=D^{2}D^{\alpha}\nabla_{\alpha}^{ab}.\label{eq:fFPoperatorSusy}
\end{equation}
This operator is the correct generalization of the FP operator as it has the right field components equations.
Therefore to see the zero mode problem  we take the eigenvalues equation
\begin{equation}
D^{2}D^{\alpha}\nabla_{\alpha}^{ab}\Lambda=\lambda\Lambda
\end{equation}
and the restriction of the functional integration to the region free of zero modes is given by the generalization of the Gribov region
\begin{equation}
\Omega:=\{\,\Gamma_{\alpha}^{b}\,|\, D^{\alpha}\Gamma_{\alpha}^{b}=0,\,\mathcal{{\cal O}}^{ab}(\Gamma_{\alpha})>0\,\}\,.\label{susyGribovregion}
\end{equation}

In order to implement the restriction to the Gribov region on the functional integration we consider the GZ approach \cite{Zwanziger1,Zwanziger2,Zwanziger25,Zwanziger3}, which consist to include a new term on the total action containing the inverse of this operator (known as horizon function), $S=S_{SYMCS}+S_{gf} + H$,  this is formally given by
\begin{align}
H(\Gamma_{\alpha}^{a})=\gamma^{4}\int d^{2}\theta\int d^{3}x\; d^{3}y\; f^{abc}\Gamma_{\alpha}^{b}(x)\left[\frac{\varepsilon^{\alpha\beta}}{D^{2}D^{\alpha}\nabla_{\alpha}}\right]^{ad}(x,y)f^{dec}\Gamma_{\beta}^{e}(y)\;.\label{hf1-1}
\end{align}
This action contain the usual Zwanziger horizon function for the gauge field in field components. This function is obtained by studying the ghost propagator as it contains the inverse of the Fadeev-Popov operator, which is invertible on the first region and diverge on the first horizon. The restriction necessary on this sector is the no pole condition \cite{Sobreiro}, which guarantees that the Faddeev-Popov operator has only positives eigenvalues. So we have the restriction implemented at the level of the gauge field.
The localization of the Gribov restriction above was done for $N=1$, $D=4$ super Yang-Mills directly in superspace in \cite{Amaral:2013uya} and in $D=3$ SYMCS in 
\cite{Amaral:2014wba}. At this point we are just interested into the gauge propagator for
 $\Gamma_{\alpha}$ and the \textit{i-particle} structure.
In order to calculate the gauge propagator we need only the bilinear of the action. Thus, for $S_{SGZ}$, we have
\begin{align}
S_{SGZ2} & =tr\int d^{3}xd^{2}\theta\Gamma_{\gamma}\frac{2\gamma^{4}}{\partial^{2}}\varepsilon^{\gamma\beta}\Gamma_{\beta}.\label{eq:supergzaction-1-1-1}
\end{align}
Following similar procedure to the SYM-CS one, the gauge propagator for SYM-CS-GZ is given by
\begin{equation}
<\Gamma_{\alpha}^{a}(1)\Gamma_{\beta}^{b}(2)>=\frac{1}{2}\delta^{ab}\left[\frac{(\partial^{4}+\gamma^{4})+im\partial^{2}D^{2}}{-(\partial^{4}+\gamma^{4})^{2}+m^{2}(\partial^{2})^{3}}\right]D^{2}D_{\beta}D_{\alpha}\delta^{2}(\theta_{1}-\theta_{2})\delta^{3}(x_{1}-x_{2}).\label{eq:SYMCSpropagatorGZ}
\end{equation}

To see how the introduction of $S_{SGZ}$ brings light on confinement
of both bosons and fermions and to compare with literature, we shall
observe the propagators in field components.
Taking components from (\ref{eq:SYMCSpropagatorGZ})
we can project the propagator for the gauge field $A_{\mu}$ 
\begin{equation}
<A_{\mu}^{a}(x_{1})A_{\nu}^{b}(x_{2})>=\delta^{ab}\left[\frac{(\partial^{4}+\gamma^{4})(-\partial^{2})}{(\partial^{4}+\gamma^{4})^{2}-m^{2}(\partial^{2})^{3}}\right](\delta_{\mu\nu}-\frac{\partial_{\mu}\partial_{\nu}}{\partial^{2}}-\frac{im\partial^{2}\varepsilon_{\mu\nu\sigma}\partial_{\sigma}}{(\partial^{4}+\gamma^{4})})\delta^{3}(x_{1}-x_{2}),
\end{equation}
and gaugino $\lambda^{\alpha}$
\begin{equation}
<\lambda_{\alpha}^{a}(x_{1})\lambda_{\beta}^{b}(x_{2})>=\frac{1}{4}\delta^{ab}\left[\frac{(\partial^{4}+\gamma^{4})}{(\partial^{4}+\gamma^{4})^{2}-m^{2}(\partial^{2})^{3}}\right](\partial^{2}\partial_{\beta\alpha}-\frac{im(\partial^{2})^{3}\varepsilon_{\beta\alpha}}{(\partial^{4}+\gamma^{4})})\delta^{3}(x_{1}-x_{2}).
\end{equation}
Despite we are interested here on the result of SYM-CS-GZ $D=3$ it is also interesting to point out the case of SYM-GZ $N=1$, $D=4$ in order to emphasize the complex pole structure. According \cite{Amaral:2013uya} the propagator
for the gauge field $a_{\mu}$ and gaugino $\lambda^{\alpha}$ are
\begin{equation}
\triangle_{a_{\mu}a_{\nu}}^{c}(1,2)=-\frac{2\partial^{2}}{\partial^{4}+\gamma^{4}}(\delta_{\mu\nu}-\frac{2\partial_{\mu}\partial_{\nu}}{\partial^{2}})\delta^{4}(x_{1}-x_{2}),
\end{equation}
\begin{equation}
\triangle_{\lambda\bar{\lambda}}^{c}(1,2)=\frac{5}{2}\frac{i\partial^{2}}{\partial^{4}+\gamma^{4}}\sigma^{\mu}\partial_{\mu}\delta^{4}(x_{1}-x_{2}).
\end{equation}

It is fundamental here to stress the structure of the poles given by the introduction of the Gribov restriction. Looking for the SYM-CS-GZ $D=3$ with $m=0$ and SYM-GZ $D=4$ is easy to observe the principal characteristic of the Gribov propagator, i.e., the complex pole structure
\begin{equation}
-\frac{2\partial^{2}}{\partial^{4}+\gamma^{4}}=\frac{1}{-\partial^{2}+i\gamma^{2}} + \frac{1}{-\partial^{2}-i\gamma^{2}}.\label{iparticlestructure}
\end{equation}
This structure is fundamental in order to exclude the gluons from the physical spectrum of the theory and also it makes possible the existence of local composite operators whose correlation functions exhibit the  K\"all\'{e}n-Lehmann spectral representation \cite{Capri:2010pg}. At this point it is worth mentioning that this property may be related to the fact that the complex pole structure break the
Osterwalder-Schrader reflection positivity condition \cite{Osterwalder:1974tc,Glimm:1979zi,Haag1992}, which states that for a particle exist as a final state it must have a well defined positive spectral density. Note that fermions are well defined as a representation of $SU(3,1)$. The Euclidean scenario require new definitions and properties and it is not unique defined but the Osterwalder-Schrader condition gives one complete definition for the Euclidean case.
So the complex pole structure makes impossible to obtain a K\"all\'{e}n-Lehmann spectral representation for a single particle. In other words, the K\"all\'{e}n-Lehmann representation of the two point correlation function is not positive. It is also important to note that is not only the Gribov propagator that breaks positivity. Many works in lattice gauge simulations indicates that positivity is broken in the infrared limit of yang-mills theory \cite{Cucchieri:2012ii,Cucchieri:2013nja,Mendes:2014gva}. Nevertheless it should be emphasized that certain combination of composite operators could admit a positive K\"all\'{e}n-Lehmann spectral representation \cite{Dudal:2010wn}. The construction of a Hilbert space for composite operators in a Gribov type model is a very difficult task and is still in initial stages of studies. 

\section{Complex N= 1, BLG-ABJM model and the Gribov ambiguity}

In the previous section we discussed the usual Gribov mechanism in $D=3$ SYMCS, which contains the gauge field sector of BLG-ABJM models \cite{Amaral:2014wba}. The complex pole structure of the \textit{i-particles}, equation (\ref{iparticlestructure}), were obtained from the superfield generalization of the Gribov-Zwanziger action. In this section we explore deeper the gauge field sector of BLG-ABJM models, in special the complex gauge model and a symmetry breaking mechanism in the scalar sector to obtain the \textit{i-particles}. Le us review the main elements of the BLG-ABJM models.
The gauge symmetry in the BLG theory is generated by a Lie $3$-algebra rather than a Lie algebra and $SO(4)$ is the only known example of a Lie $3$-algebra. It is possible to decompose the gauge symmetry generated by $SO(4)$ into $SU(2) \times SU(2)$. In these way it is possible to write the gauge symmetry of BLG as generated by ordinary Lie algebras and the gauge sector of the theory is now given by two Chern-Simons cocycles with levels $\pm k$ and the 
matter fields  exist in the so-called bi-fundamental representation. The BLG theory represents two M2-branes due to the fact that its gauge symmetry  is generated by the gauge group 
 $SU(2)_k \times SU(2)_{-k}$. However, it has been possible to extend the gauge group 
 to  $U(N)_k \times U(N)_{-k}$, and the resultant theory is called ABJM 
theory. Let us present the typical SCS gauge sector of two $SU(2)$ fields as it appears in a BLG-ABJM model:
\begin{equation}
S_{SCS}=i\int d^{3}xd^{2}\theta\left[(D^{\alpha}\Gamma^{a\beta})(D_{\beta}\Gamma_{\alpha}^{a})+\frac{2}{3}igf^{abc}\Gamma^{a\alpha}\Gamma^{b\beta}(D_{\beta}\Gamma_{\alpha}^{c})-\frac{1}{6}g^{2}f^{abc}f^{cde}\Gamma^{a\alpha}\Gamma^{b\beta}\Gamma_{\alpha}^{d}\Gamma_{\beta}^{e}\right].\label{eq:actionSCS-1abjm}
\end{equation}
As we are interested in N = 1 supersymmetric gauge field theory with
the gauge group $G\times G$, we write a second action for another
gauge superfield $\tilde{\Gamma}_{\alpha}^{a}$ 
\begin{equation}
\tilde{S}_{SCS}=i\int d^{3}xd^{2}\theta\left[(D^{\alpha}\tilde{\Gamma}^{a\beta})(D_{\beta}\tilde{\Gamma}_{\alpha}^{a})+\frac{2}{3}igf^{abc}\tilde{\Gamma}^{a\alpha}\tilde{\Gamma}^{b\beta}(D_{\beta}\tilde{\Gamma}_{\alpha}^{c})-\frac{1}{6}g^{2}f^{abc}f^{cde}\tilde{\Gamma}^{a\alpha}\tilde{\Gamma}^{b\beta}\tilde{\Gamma}_{\alpha}^{d}\tilde{\Gamma}_{\beta}^{e}\right].\label{eq:actionSCS-2abjm}
\end{equation}
And
 \begin{equation}
S_{BLG}(\Gamma,\tilde{\Gamma})=S_{SCS}(\Gamma)-\tilde{S}_{SCS}(\tilde{\Gamma}).\label{eq:actionABJM}
\end{equation}

It is interesting to note that the bilinear sector of these action  
\begin{equation}
S^{0}(\Gamma,\tilde{\Gamma}) =i\int d^{3}xd^{2}\theta\left[(D^{\alpha}\Gamma^{a\beta})(D_{\beta}\Gamma_{\alpha}^{a}) - (D^{\alpha}\tilde{\Gamma}^{a\beta})(D_{\beta}\tilde{\Gamma}_{\alpha}^{a})\right],
\end{equation}
can be written as a complex CS in a holomorphic form using two complex fields
\begin{align}
 I_{\beta}^{a} & = \frac{1}{\sqrt{2}}(\Gamma_{\beta}^{a}+i \tilde{\Gamma}_{\beta}^{a})\nonumber \\
 I_{\beta}^{a\dagger} & = \frac{1}{\sqrt{2}}(\Gamma_{\beta}^{a}-i \tilde{\Gamma}_{\beta}^{a}),
\end{align}
in a way that
\begin{align}
S^{0}(I,I^{\dagger}) & =i\int d^{3}xd^{2}\theta\left[(D^{\alpha}I^{a\beta})(D_{\beta}I_{\alpha}^{a}) + (D^{\alpha}I^{a\dagger\beta})(D_{\beta}I_{\alpha}^{a\dagger})\right],\nonumber \\
S^{0}(\Gamma,\tilde{\Gamma}) & = S^{0}_{ABJM}(I,I^{\dagger}).
\end{align}
In fact the construction of a complex CS action is not new and Witten \cite{Witten:2010cx,Witten:2010zr} uses Morse theory in order to give consistent mathematical rigor to a complex CS theory and define suitable integration contours in a Feynmann functional integral of a CS action with complex field. In other terms, Witten was able to define integration contours for two complex CS in a holomorphic construction.

The $S^{0}(I,I^{\dagger})$ corresponds to a $U(1)_k \times U(1)_{-k}$ and is clearly a holomorphic construction or in more simple terms a real abelian action that can be constructed with two complex fields that have complex gauge transformations i.e,
\begin{align}
\delta I_{\alpha}^{a} & =(D_{\alpha}(I)\Lambda)^{a},\nonumber \\
\delta I_{\alpha}^{a\dagger} & =(D_{\alpha}(I)\Lambda)^{a\dagger}.
\label{eq:complexinv}
\end{align}

Before proceeding to build a complex version of the $SU(2)_k \times SU(2)_{-k}$ it is interesting to remember that the Gribov type correlator forbids a K\"all\'{e}n-Lehmann spectral representation for a single particle but simultaneously, due to the complex conjugate structure of the pole, it is possible to obtain composite operators that admits a positive K\"all\'{e}n-Lehmann spectral representation. This is the fundamental point that connects the complex holomorphic action to the confinement in the Gribov scenario. 
Now the generalization from the complex $U(1)_k \times U(1)_{-k}$ to the complex $SU(2)_k \times SU(2)_{-k}$ is straightforward
\begin{align}
\delta I_{\alpha}^{a} & =(\nabla_{\alpha}(I)\Lambda)^{a}\nonumber \\
\nabla_{\alpha}^{ab} & =\delta^{ab}D_{\alpha}+gf^{acb}I_{\alpha}^{c}
\label{eq:complexinv}
\end{align}
and the complex conjugate 
\begin{align}
\delta I_{\alpha}^{a\dagger} & =(\nabla_{\alpha}(I)\Lambda)^{a\dagger}\nonumber \\
\nabla_{\alpha}^{ab\dagger} & =\delta^{ab}D_{\alpha}+gf^{acb}I_{\alpha}^{c\dagger}.
\label{eq:complexinv2}
\end{align}
The combination 
\begin{eqnarray}
S_{ABJM} &=& i\int d^{3}xd^{2}\theta\left[(D^{\alpha}I^{a\beta})(D_{\beta}I_{\alpha}^{a})+\frac{2}{3}igf^{abc}I^{a\alpha}I^{b\beta}(D_{\beta}I_{\alpha}^{c})-\frac{1}{6}g^{2}f^{abc}f^{cde}I^{a\alpha}I^{b\beta}I_{\alpha}^{d}I_{\beta}^{e}\right]\nonumber \\
&+& i \int d^{3}xd^{2}\theta\left[(D^{\alpha}I^{\dagger a\beta})(D_{\beta}I_{\alpha}^{\dagger a})+\frac{2}{3}igf^{abc}I^{\dagger a\alpha}I^{\dagger b\beta}(D_{\beta}I_{\alpha}^{\dagger c})-\frac{1}{6}g^{2}f^{abc}f^{cde}I^{\dagger a\alpha}I^{\dagger b\beta}I_{\alpha}^{\dagger d}I_{\beta}^{\dagger e}\right] \nonumber \\
S_{ABJM}&=&S_{SCS}(I)+S_{SCS}(I^{\dagger}),\label{eq:actionABJM-complex}
\end{eqnarray}
is gauge invariant, real and the bilinear sector is exactly the same as presented in (\ref{eq:actionABJM}), and also has holomorphic independent sectors.

If we compute the propagators for gauge fields from the action (\ref{eq:actionABJM-complex}) we don't have the complex poles \textit{i-particles} structure. We propose an alternative to the Gribov mechanism, which involves the introduction of a scale. It is important to remember that in the BLG model there are matter fields in the bifundamental representation. In these way, it is interesting to construct the matter sector in terms of complex fields and use a symmetry breaking mechanism in order to obtain a mass for the complex gauge fields \cite{Mukhi2008,Mauri2008,Mukhi2011}. An appropriated symmetry breaking mechanism can provide the \textit{i-particles} structure \cite{Guimaraes2016,Capri2013,Vilar2011}.
So let us consider two complex scalar fields, in the adjoint representation, $\varphi^{a}$ and $\varphi^{a\dagger}$ that transforms according
\begin{eqnarray}
\delta\varphi^{a} &=& gf^{acb}\Lambda^{c}\varphi^{b} \nonumber \\
\delta\varphi^{\dagger} &=& gf^{acb}\Lambda^{c\dagger}\varphi^{b\dagger},
\end{eqnarray}
and the two complex covariant derivatives
\begin{eqnarray}
(\nabla_{\alpha}\varphi)^{a} &=& \nabla_{\alpha}^{ab}\varphi^{b} \nonumber \\
(\nabla_{\alpha}\varphi )^{a\dagger} &=& \nabla_{\alpha}^{ab\dagger}\varphi^{b\dagger}, 
\end{eqnarray}
that transforms as
\begin{eqnarray}
\delta (\nabla_{\alpha}\varphi)^{a} &=& gf^{acb}\Lambda^{c}(\nabla_{\alpha}\varphi)^{b} \nonumber \\
\delta (\nabla_{\alpha}\varphi)^{a\dagger} &=& gf^{acb}\Lambda^{c\dagger}(\nabla_{\alpha}\varphi)^{b\dagger}.
\end{eqnarray}
%
In order to stay as close as possible to the BLG-ABJM case we will limit our potential to a quartic one that admits a non zero expectation value for the scalar field.
An interesting case to consider for the matter action is given by
\begin{equation}
S_{mix}=\int d^{3}xd^{2}\theta ( a(\nabla^{\beta}\varphi)^{a}(\nabla_{\beta}\varphi)^{a} + a^{\dagger}(\nabla^{\beta}\varphi )^{a\dagger}(\nabla_{\beta}\varphi )^{a\dagger}  
-\frac{\lambda}{4}(\varphi^{a}\varphi^{a}-\varphi^{a\dagger}\varphi^{a\dagger})^{2} ) \label{matteractionblg}
\end{equation}
and we will consider now $a=1+i$ and $a^{\dagger}=1-i$. This case certainly do not corresponds to real mass poles but surely to complex mass poles as it appear in Gribov correlator. This fact is related to the breaking of the Osterwald Schrader positivity condition. The breaking of positivity is fundamental in order to ensure that there is no single particle state. The invariant action is given by
\begin{equation}
S_{inv}= S_{ABJM}+S_{mix}.
\end{equation}
Now it is important to remember that due to the gauge symmetry it is necessary to perform a gauge fixing and following the Faddeev-Popov procedure and taking the Landau gauge for the two complex gauge fields we end with the gauge fixing action
\begin{equation}
S_{gf}=\frac{1}{4}s\{\int d^{3}xd^{2}\theta(c'^{a}D^{\alpha}(I_{\alpha}^{a}-i I_{\alpha}^{\dagger a}) + c'^{\dagger a}D^{\alpha}(I_{\alpha}^{\dagger a}+iI_{\alpha}^{a}))\}, 
\end{equation}
where the Faddeev-Popov fields $c'^{a}$ and $c^{a}$ are the antighost and the ghost respectively.
And $s$ is the BRST nilpotent operator ($s^{2}=0)$. It is Clear that this structure is constructed not only to write a BRST symmetry but also to define a real Gribov operator and permits the definition of the first Gribov region. It is clear that this gauge fixing, when written in terms of real fields offers the same property of the usual Landau gauge used in SYM-CS presented in the previous section.
The total action $S=S_{inv}+S_{gf}$ is invariant under
\begin{align}
sI_{\alpha}^{a} & =(\nabla_{\alpha}c)^{a}\hspace{2cm} sI_{\alpha}^{\dagger a} =(\nabla_{\alpha}c)^{\dagger a}\nonumber \\
sc^{a} & =-\frac{1}{2}gf_{abc}c^{b}c^{c}\hspace{1,2cm} sc^{\dagger a} =-\frac{1}{2}gf_{abc}c^{\dagger b}c^{\dagger c}\nonumber \\
sc'^{a} & =b^{a}\hspace{2,7cm} sc'^{\dagger a}  =b^{\dagger a}\nonumber \\
sb^{a} & =0,\hspace{2,8cm} sb^{\dagger a}  =0 \nonumber \\
s\varphi^{a} & = gf^{acb}c^{c}\varphi^{b} \nonumber \\
s\varphi^{\dagger} & = gf^{acb}c^{c\dagger}\varphi^{b\dagger}
\label{eq:brst-I}.
\end{align}
This gauge fixing can be written in terms of the original real fields $\Gamma,\tilde{\Gamma}$ and, as in the case of SYM-CS, in order to stay at the first Gribov region, a necessary and sufficient condition is that the real gauge fields has a Gribov type correlator. Now it is necessary, in order to stay at the first Gribov region that the correlator for the complex fields corresponds to \textit{i-particles}. After a symmetry breaking\footnote{It is always important to remember that in a complex theory like the one that we are discussing the concept of a symmetry breaking must be understood in the context of consistent integration contours in a Feynmann functional integral and can be taken just to access consistently the infrared sector of the model.}, and taking the expectation value of the scalar field in diagonal direction, we have for the bilinear 
\begin{align}
 & \int d^{3}xd^{2}\theta\left[i(D^{\alpha}I^{j\beta})(D_{\beta}I_{\alpha}^{j}) +i(D^{\alpha}I^{j\dagger\beta})(D_{\beta}I_{\alpha}^{j\dagger})+\gamma\left(a I^{j\alpha}I_{\alpha}^{j}+ a^{\dagger}I^{j\dagger\alpha}I_{\alpha}^{j\dagger}\right)\right]\nonumber \\
= & \int d^{3}xd^{2}\theta\left[i(I_{\beta}^{j}D^{\alpha}D^{\beta}I_{\alpha}^{j})
+i(I_{\beta}^{j\dagger}D^{\alpha}D^{\beta}I_{\alpha}^{j\dagger})+\gamma\left(a\varepsilon^{\alpha\beta}I_{\beta}^{j}I_{\alpha}^{j}+a^{\dagger}\varepsilon^{\alpha\beta}I_{\beta}^{j\dagger}I_{\alpha}^{j\dagger}\right)\right],
\end{align}
Where $j$ represents the sum over the non-diagonal directions.
Remembering that $a$ and $a^{\dagger}$ are complex and $\gamma$ is the expectation value for two scalar fields, $\gamma=<\varphi^{a}\varphi^{a}>g^{2}$. Note that we are taking the expectation vale for the scalar field in the diagonal direction and we are considering the calculation for the $SU(2)$ group. The definition of the covariant derivatives in (\ref{eq:complexinv}, \ref{eq:complexinv2}) turns this calculation straightforward. It is easy to see that the case with $a=1$ and $a^{\dagger}=1$ corresponds to two real mass poles. The most general case of a holomorphic action is given by $a=1-i$ and $a^{\dagger}=1+i$ and the obtained correlator corresponds to a Gribov Zwanziger type one. The correlators in momentum space are given by
\begin{eqnarray}
<I_{\alpha}^{j}(1)I_{\beta}^{k}(2)> & = & \frac{1}{4}\delta^{jk}\left[ \frac{2iD^{2}+(1-i)\gamma}{p^{2}-i\frac{\gamma^{2}}{2}}\right] \frac{D^{2}}{p^{2}}D_{\beta}D_{\alpha}\delta^{2}(\theta_{1}-\theta_{2})\delta^{3}(x_{1}-x_{2}),\nonumber \\
<I_{\alpha}^{j\dagger}(1)I_{\beta}^{k\dagger}(2)> & = & \frac{1}{4}\delta^{jk}\left[ \frac{2iD^{2}+(1+i)\gamma}{p^{2}+i\frac{\gamma^{2}}{2}}\right] \frac{D^{2}}{p^{2}}D_{\beta}D_{\alpha}\delta^{2}(\theta_{1}-\theta_{2})\delta^{3}(x_{1}-x_{2}). \label{ipartpropagforI}
\end{eqnarray}
It is easy to note that the propagator in terms of the real fields $\Gamma,\tilde{\Gamma}$ is real and is in the first Gribov region. In fact in terms of real fields is straightforward to obtain the propagators as
\begin{eqnarray}
<\Gamma_{\alpha}^{j}(1)\Gamma_{\beta}^{k}(2)> & = &\frac{1}{4}\delta^{jk}\left[ \frac{2iD^{2}p^{2}+\gamma(p^{2}+\frac{\gamma^{2}}{2})}{p^{4}+\frac{\gamma^{4}}{4}}\right] \frac{D^{2}}{p^{2}}D_{\beta}D_{\alpha}\delta^{2}(\theta_{1}-\theta_{2})\delta^{3}(x_{1}-x_{2}),\nonumber \\ 
<\Gamma_{\alpha}^{j}(1)\tilde{\Gamma}_{\beta}^{k}(2)> & = &-\frac{1}{4}\delta^{jk}\left[ \frac{iD^{2}\gamma^{2}+\gamma(p^{2}-\frac{\gamma^{2}}{2})}{p^{4}+\frac{\gamma^{4}}{4}}\right] \frac{D^{2}}{p^{2}}D_{\beta}D_{\alpha}\delta^{2}(\theta_{1}-\theta_{2})\delta^{3}(x_{1}-x_{2}),\nonumber \\
<\tilde{\Gamma}_{\alpha}^{j}(1)\tilde{\Gamma}_{\beta}^{k}(2)> & = & - <\Gamma_{\alpha}^{j}(1)\Gamma_{\beta}^{k}(2)>.
\end{eqnarray}
It is important to emphasize here that these propagators are used to calculate the two point ghost function and the Gribov pole ensure that we are in the first Gribov region due to the fact that the two point ghost function goes to infinity at the Gribov frontier, which is a well known property that indicate that we have the correct restriction \cite{Guimaraes2016,Capri2013,Sobreiro}. This calculation in $N=1$ is done in \cite{Amaral:2014wba} for real gauge field and is straightforward to see that those propagators imply into the first Gribov region.

It now remains to obtain a candidate to composite operator that exhibit the K\"all\'{e}n-Lehmann spectral representation. That operator must obey the definition in 3 dimensions
\begin{equation}
\langle O(k) O(-k)\rangle \rightarrow {\cal I}(k^2,m_{1},m_{2}) =\int \frac{d^3 p}{(2\pi)^3}\frac{F(p,k-p)}{\left( (k-p)^2+m_{1}^2 \right) \left(  p^2 + m_{2}^2 \right)}
\end{equation}
and ${\cal I}(k^2,m_{1},m_{2})$ can be written as \cite{Dudal:2010wn}
\begin{equation}
 {\cal I}(k^2,m_{1},m_{2})= \int_{\mu^2}^{\infty} d\tau \rho(\tau) \left( \frac{1}{\tau+k^2} \right)  \;, \label{iint} 
\end{equation}
with a positive spectral density. So, ${\cal I}(k^2,m_{1},m_{2})$ represents a propagator of a composite particle.

It is well known that the Gribov correlator only admits the K\"all\'{e}n-Lehmann spectral representation if a product of two conjugate \textit{i-particle} appears simultaneously. Using the results obtained in \cite{Dudal:2010wn}, the condensate made of two \textit{i-particle} correlators ${\cal I}(k^2)$ is of the form
\begin{equation}
{\cal I}(k^2) = \int \frac{d^3p}{(2\pi)^3} \; \frac{1}{\left( (k-p)^2+i\sqrt{2}\vartheta^2 \right) \left(  p^2 -i \sqrt{2}\vartheta^2 \right)}  \;,  \label{bint}
\end{equation}
which can be written as  
\begin{equation}
{\cal I}(k^2) = \int_{\tau_{0} }^{\infty }d\tau \frac{\rho(\tau )}{\tau +k^{2}}  \;, \label{iint}
\end{equation}
with  
\begin{equation}
  \tau_{0} \leq \tau \leq  \infty ,
\end{equation}
and the spectral density 
\begin{equation}
\rho(\tau) = \frac{1}{8\pi} \frac{1}{\sqrt{\tau}}  \;, \label{spectr}
\end{equation}
which is positive throughout the integration range and ensure that a two \textit{i-particles} composite operator admits a K\"all\'{e}n-Lehmann spectral representation and this operator can be seen as a massive particle. If a composite operator $\langle {\cal O}(k) {\cal O}(-k) \rangle$ can be cast in the form (\ref{iint}) then there is a condensation that allows to difine it as corresponding to a state of a composite particle \cite{Baulieu:2009ha}. Note that in (\ref{ipartpropagforI}) we obtained propagators with poles $p^{2}+i\frac{\gamma^{2}}{2}$ and $p^{2}-i\frac{\gamma^{2}}{2}$, which give the Gribov structure when combined as in (\ref{iparticlestructure}). It is also true that due to the holomorphic structure of the action there is no mix between the two complex correlators and even taking high order loop corrections the spectral representation is always sustained. So, we need a composite operator that combines the propagators for $I_{\alpha}^{j}$, $I_{\alpha}^{i\dagger}$ in (\ref{ipartpropagforI}) to give the pole structure $(p^{2}+i\frac{\gamma^{2}}{2})(p^{2}-i\frac{\gamma^{2}}{2})$.
A natural candidate for a composite operator in these terms is of the form
\begin{equation}
 {\cal O}=\int d^{2}\theta\left[i D^{\gamma}D^{\beta}I_{\gamma}^{j}D^{\alpha}D_{\beta}I_{\alpha}^{j\dagger}\right] 
\end{equation}
which its two point correlation function can be cast in the form of a spectral representation with positive spectral function as 
given in (\ref{iint}), namely:
\begin{eqnarray}
\langle {\cal O}(k) {\cal O}(-k) \rangle & = & \int_{\tau_{0}}^{\infty} d\tau \frac{\rho(\tau)}{\tau+k^2}. \label{opcomp}
\end{eqnarray}
This operator is not gauge invariant but the composite operator constructed from the full operator (\ref{woperator}) is a candidate that offers a gauge invariant composite particle and is given by
\begin{equation}
 {\cal O}=\int d^{3}x d^{2}\theta \left\lbrace W^{a\alpha}(I)W_{\alpha}^{a}(I^{\dagger})\right\rbrace .
\end{equation}

So we have provided the (first order) contribution to the two-point function  of the gauge invariant condensate in the ABJM framework getting the same structure of the usual Gribov mechanism, which can be an interesting start point to obtain glueball masses \cite{Capri2011}.
It is important to stress here that due to the complex holomorphic structure of the action this properties could be extended beyond the one loop level here considered and take into account loop corrections. It is also important to note that not all operators constructed with the complex fields are acceptable physical operators. For example the gauge sector of the action does not offer the same type of structure if we take into account a Wick prescription as defined in \cite{Dudal:2010wn}. Only a mixing with equal numbers of $I$ and $I^{\dagger}$ has the possibility to be associated to an observable. The study of all kind of observables with these type of structure will be done in a future work.

\section{Conclusions}

In this work we have studied some aspects of a simplified version of BLG-ABJM kind of models using SCS action structure. The complex pole structure of the Gribov propagator and the possibility of obtaining this pole structure from a complex holomorphic action was presented. Also it was discussed the a general complex scalar gauge invariant holomorphic action. Due to a symmetry breaking mechanism for this complex scalar action the characteristic Gribov pole structure is obtained in the case of a general kinetic holomorphic cocycle of the scalar action, equation (\ref{matteractionblg}). Also a good candidate to a composite operator in order to generate a K\"all\'{e}n-Lehmann spectral representation, that has positive spectral density, which characterize a composite particle state, was obtained. It is interesting to note that the holomorphic structure is fundamental in order to introduce interaction terms in the definition of the composite operator and perform loop corrections without a mix between correlators that destroy the spectral representation. This mechanism could be fundamental in the study of confinement in the Gribov scenario and will be matter for a future work.
To conclude we can point out for the necessity of absorbing Witten's work on holomorphic complex theories \cite{Witten:2010cx} in order to settle the physical interpretation of this non-perturbative scenario, that is, theories that provide propagators with the complex pole structure of the \textit{i-particles}.

\section*{Acknowledgments}

The Conselho Nacional de Desenvolvimento Cient\'{\i}fico e tecnol\'{o}gico
CNPq- Brazil, Funda\c{c}\~{a}o de Amparo a Pesquisa do Estado do Rio de Janeiro
(Faperj), the SR2-UERJ and the Coordena\c{c}\~{a}o de Aperfeioamento de Pessoal
de N\'{\i}vel Superior (CAPES) are acknowledged for the financial support. 

\appendix
\section[Appendix]{Notation, conventions and some useful formulas}

We work with Euclidean metric: diag(+++). So we choose the gamma matrices
being the Pauli matrices $\sigma_{i}$ \cite{GradedMajorana}:

\begin{equation}
\gamma^{\mu}\equiv(\sigma_{\mu})_{\alpha}^{\:\beta}
\end{equation}
witch are Osterwalder-Schrader (OS) self-conjugate and
\begin{equation}
\{\sigma^{\mu},\sigma^{\nu}\}=2\delta^{\mu\nu}I,
\end{equation}
\begin{equation}
[\sigma^{\mu},\sigma^{\nu}]=2i\varepsilon^{\mu\nu\sigma}\sigma^{\sigma}.
\end{equation}
The invariant anti-symmetric tensor is defined as
\begin{equation}
\varepsilon^{-+}=\varepsilon_{-+}=+1,
\end{equation}
\begin{equation}
\varepsilon^{\gamma\beta}\varepsilon_{\beta\alpha}=-\delta_{\alpha}^{\gamma},\label{eq:epsoncontraction}
\end{equation}
and are used to raise and lower indices
\begin{equation}
\psi^{\alpha}=\varepsilon^{\alpha\beta}\psi_{\beta},
\end{equation}
\begin{equation}
\psi_{\alpha}=\psi^{\beta}\varepsilon_{\beta\alpha}.
\end{equation}

The representation of differential
operator of the generators of super algebra in D=3 is described with the concept
of graded Majorana \cite{GradedMajorana}
\begin{equation}
Q_{\alpha}=-\partial_{\alpha}+\partial_{\alpha\beta}\theta^{\beta},
\end{equation}
with
\begin{equation}
\partial_{\alpha\beta}=i\sigma_{\alpha}^{\mu\gamma}\varepsilon_{\gamma\beta}\partial_{\mu}.
\end{equation}
As well as the superspace derivative
\begin{equation}
D_{\alpha}=\partial_{\alpha}+\partial_{\alpha\beta}\theta^{\beta},
\end{equation}
with the following relations
\begin{equation}
\{D_{\alpha},D_{\beta}\}=2\partial_{\alpha\beta},
\end{equation}
\begin{equation}
[D_{\alpha},D_{\beta}]=-2\varepsilon_{\alpha\beta}D^{2},
\end{equation}
\begin{equation}
D_{\alpha}D_{\beta}=\partial_{\alpha\beta}-\varepsilon_{\alpha\beta}D^{2},\label{eq:superderivativealfabetarelation}
\end{equation}
\begin{equation}
D^{\beta}D_{\alpha}D_{\beta}=0.\label{eq:superderivativealfabetaalfarelationzero}
\end{equation}
\begin{equation}
[Q_{\alpha},D_{\beta}]=0.
\end{equation}

Another useful relations:

\begin{equation}
\partial_{\alpha\beta}\partial^{\alpha\gamma}=\partial^{2}\delta_{\beta}^{\gamma},\label{eq:realtionderivativecontraction}
\end{equation}
\begin{equation}
(D^{2})^{2}=-\partial^{2},\label{eq:D2squared}
\end{equation}
\begin{equation}
\int d^{2}\theta=-\frac{1}{4}D^{2}.
\end{equation}


\begin{thebibliography}{10}



\bibitem{Li1995}
 	M. Li, C-I. Tan, 
 	Phys. Rev. D 51, 328, (1995).

\bibitem{Ivanov1998}
	 D. Y. Ivanov, R. Kirschner, E. M. Levin, L. N. Lipatov, L. Szymanowski, M. Wusthoff, 
	Physical Review D, 58(7), (1998). 

\bibitem{Mathieu:2008me} 
  V.~Mathieu, N.~Kochelev and V.~Vento,
  Int.\ J.\ Mod.\ Phys.\ E {\bf 18}, 1 (2009).
  [arXiv:0810.4453 [hep-ph]].
  
\bibitem{Witten1989}
	Edward Witten
	Commun. Math. Phys. 121, 351-399 (1989).

\bibitem{Witten:2010cx} 
  E.~Witten,
  arXiv:1001.2933 [hep-th].

\bibitem{Witten:2010zr} 
  E.~Witten,
  arXiv:1009.6032 [hep-th].
  
\bibitem{Gribov}V.~N.~Gribov,
  Nucl.\ Phys.\ B {\bf 139}, 1 (1978).

\bibitem{Zwanziger1}D.~Zwanziger,
  Nucl.\ Phys.\ B {\bf 321}, 591 (1989).

\bibitem{Zwanziger2}D.~Zwanziger,
  Nucl.\ Phys.\ B {\bf 323}, 513 (1989).

\bibitem{Zwanziger25}D.~Zwanziger,
  Nucl.\ Phys.\ B {\bf 378}, 525 (1992).

\bibitem{Zwanziger3}D.~Zwanziger,
  Nucl.\ Phys.\ B {\bf 399}, 477 (1993).

\bibitem{rgzmodel}D.~Dudal, J.~A.~Gracey, S.~P.~Sorella, N.~Vandersickel and H.~Verschelde,
  Phys.\ Rev.\ D {\bf 78}, 065047 (2008)
  [arXiv:0806.4348 [hep-th]].

\bibitem{Sorella:2010it} 
  S.~P.~Sorella,
  J.\ Phys.\ A {\bf 44}, 135403 (2011)
  [arXiv:1006.4500 [hep-th]].

\bibitem{Capri:2010pg} 
  M.~A.~L.~Capri, A.~J.~Gomez, M.~S.~Guimaraes, V.~E.~R.~Lemes, S.~P.~Sorella and D.~G.~Tedesco,
  Eur.\ Phys.\ J.\ C {\bf 71}, 1525 (2011)
  [arXiv:1009.3062 [hep-th]].

\bibitem{singer}
I. M. Singer.
\newblock Some Remarks on the Gribov Ambiguity. 
\newblock {\em Commun. Math. Phys.} 60, 7 (1978)

\bibitem{Baal1992}P.~van Baal,
  Nucl.\ Phys.\ B {\bf 369}, 259 (1992).
  
\bibitem{labast}E.~Witten,
  J.\ Diff.\ Geom.\  {\bf 17}, 661 (1982).\newline
  
J.~M.~F.~Labastida,
  Commun.\ Math.\ Phys.\  {\bf 123}, 641 (1989).

\bibitem{milnor}J. Milnor, Ann. Math. Stud. 51, Princeton Univ. Press, Princeton NJ (1973).

\bibitem{matsumoto}Matsumoto, Yukio, ``An Introduction to Morse Theory'', Iwanami series in modern mathematics {\bf 208}, AMS Bookstore (2002).


\bibitem{Guimaraes2016}
	M. S. Guimaraes, A. D. Pereira, S. P. Sorella,
	Phys. Rev. D 94, 116011 (2016).

\bibitem{Capri2013}
	M. A. L. Capri, D. Dudal, M. S. Guimaraes, L. F. Palhares, S. P. Sorella,
	Phys. Lett. B. 01, 039, (2013).

\bibitem{Vilar2011}
	L. C. Q. Vilar, O.S. Ventura, V. E. R. Lemes,
	Phys.\ Rev.\ D {\bf 84}, 105026 (2011).
  
\bibitem{Bagger}J.~Bagger and N.~Lambert,
  Phys.\ Rev.\ D {\bf 75}, 045020 (2007)
  [hep-th/0611108].
  
J.~Bagger and N.~Lambert,
  Phys.\ Rev.\ D {\bf 77}, 065008 (2008)
  [arXiv:0711.0955 [hep-th]].
  
J.~Bagger and N.~Lambert,
  JHEP {\bf 0802}, 105 (2008)
  [arXiv:0712.3738 [hep-th]].
  
\bibitem{Gustavsson:2007vu} 
  A.~Gustavsson,
  Nucl.\ Phys.\ B {\bf 811}, 66 (2009)
  [arXiv:0709.1260 [hep-th]].
  
\bibitem{Aharony:2008ug} 
  O.~Aharony, O.~Bergman, D.~L.~Jafferis and J.~Maldacena,
  JHEP {\bf 0810}, 091 (2008)
  [arXiv:0806.1218 [hep-th]].



\bibitem{Brower2000}
	R. C. Brower, S. D. Mathur, C-I. Tan,
	Nucl. Phys. B 587:249-276, (2000).

  
\bibitem{Colangelo2007}
	P. Colangelo, F. De Fazio, F. Jugeau, S. Nicotri,
 	Phys. Lett. B 652:73-78, (2007).

\bibitem{Mukhi2008}
	S. Mukhi, C. Papageorgakis,
	JHEP 0805:085, (2008).
	[arXiv:0803.3218 [hep-ph]].

\bibitem{Mauri2008}
	A. Mauri, A. C. Petkou,
	Phys. Lett. B 666:527-532, (2008).
	[arXiv:0806.2270 [hep-ph]].

\bibitem{Mukhi2011}
	S. Mukhi,
	JHEP 12 083, (2011).
	[arXiv:1110.3048 [hep-ph]].



\bibitem{Kugo:1982bn}T.~Kugo and P.~K.~Townsend,
  Nucl.\ Phys.\ B {\bf 221}, 357 (1983).

\bibitem{McKeon:2000qm} 
  D.~G.~C.~McKeon,
  Nucl.\ Phys.\ B {\bf 591}, 591 (2000).

\bibitem{McKeon:2001pm} 
  D.~G.~C.~McKeon and T.~N.~Sherry,
  Annals Phys.\  {\bf 288}, 2 (2001).

\bibitem{Amaral:2013uya} 
  M.~M.~Amaral, Y.~E.~Chifarelli and V.~E.~R.~Lemes,
  J. Phys. A: Math. Theor. 47 (2014) 075401.
  hep-th/1310.8250.

\bibitem{Wetterich:2010ni} 
  C.~Wetterich,
  Nucl.\ Phys.\ B {\bf 852}, 174 (2011)
  [arXiv:1002.3556 [hep-th]].

\bibitem{RuizRuiz1} 
  F.~Ruiz Ruiz and P.~van Nieuwenhuizen,
  Nucl.\ Phys.\ Proc.\ Suppl.\  {\bf 56B}, 269 (1997)
  [hep-th/9701052].

\bibitem{Sobreiro} 
	R.F. Sobreiro, S.P. Sorella.
	[hep-th/0504095].

\bibitem{Amaral:2014wba} 
  M.~M.~Amaral and V.~E.~R.~Lemes,
  Annals Phys.\  {\bf 356}, 505 (2015)
  [arXiv:1410.6768 [hep-th]].

\bibitem{Osterwalder:1974tc} 
  K.~Osterwalder and R.~Schrader,
  Commun.\ Math.\ Phys.\  {\bf 42}, 281 (1975).

\bibitem{Glimm:1979zi} 
  J.~Glimm and A.~M.~Jaffe,
  Lett.\ Math.\ Phys.\  {\bf 3}, 377 (1979).

\bibitem{Haag1992} 
  R.~Haag,
  ``Local quantum physics: Fields, particles, algebras''.
  Berlin, Germany: Springer, 356 p. (1992).


\bibitem{Cucchieri:2012ii} 
  A.~Cucchieri and T.~Mendes,
  Phys.\ Rev.\ D {\bf 86}, 071503 (2012)
  [arXiv:1204.0216 [hep-lat]].

v\bibitem{Cucchieri:2013nja} 
  A.~Cucchieri and T.~Mendes,
  Phys.\ Rev.\ D {\bf 88}, 114501 (2013)
  [arXiv:1308.1283 [hep-lat]].

\bibitem{Mendes:2014gva} 
  T.~Mendes and A.~Cucchieri,
  PoS LATTICE {\bf 2013}, 456 (2014)
  [arXiv:1401.6908 [hep-lat]].

\bibitem{Dudal:2010wn} 
  D.~Dudal and M.~S.~Guimaraes,
  Phys.\ Rev.\ D {\bf 83}, 045013 (2011)
  [arXiv:1012.1440 [hep-th]].

\bibitem{Baulieu:2009ha}
L.~Baulieu, D.~Dudal, M.~Guimaraes, M.~Huber, S.~Sorella, N.~Vandersickel and D.~Zwanziger,
Phys. Rev. D \textbf{82}, 025021 (2010)
[arXiv:0912.5153 [hep-th]].

\bibitem{Capri2011}
	M. A. L. Capri, A. J. Gomez, M. S. Guimaraes, V. E. R. Lemes, S. P. Sorella, D. G. Tedesco,
	Phys. Rev. D. 85, 085012, (2011).

\bibitem{GradedMajorana}A.~F.~Schunck and C.~Wainwright,
  J.\ Phys.\ A {\bf 39}, 3787 (2006)
  [hep-th/0501252].
  


\end{thebibliography}
\end{document}